\documentclass[sigconf]{acmart}

\usepackage{soul}
\usepackage{graphicx} 
\usepackage{float}
\copyrightyear{2024}
\acmYear{2024}
\setcopyright{rightsretained}
\acmConference[AST '24]{5th ACM/IEEE International Conference on Automation of Software Test (AST 2024)}{April 15--16, 2024}{Lisbon, Portugal}
\acmBooktitle{5th ACM/IEEE International Conference on Automation of Software Test (AST 2024) (AST '24), April 15--16, 2024, Lisbon, Portugal}\acmDOI{10.1145/3644032.3644457}
\acmISBN{979-8-4007-0588-5/24/04}
\usepackage{hyperref}
\href{}{}

\usepackage[most]{tcolorbox}
\NewTColorBox{custombox}{o}{
    colback=gray!10,
	colframe=gray!10,
	left=1.0mm,
	right=1.0mm,
	top=1.0mm,
	bottom=1.0mm,
	fonttitle=\bfseries,
	arc=0mm,
	leftrule=0mm,
	rightrule=0mm,
	toprule=0mm,
	bottomrule=0mm,
	notitle,
	before=\par\medskip\noindent,
    IfValueTF={#1}{before upper={\textbf{#1: } }}{},
    parbox=false,
}

\title{PAFOT: A Position-Based Approach for Finding Optimal Tests of Autonomous Vehicles}

\author{Victor Crespo-Rodriguez}
\email{victor.cresporodriguez@monash.edu}
\orcid{0000-0003-2783-098X}
\affiliation{%
\institution{Faculty of Information Technology,}
  \institution{Monash University}
  \city{Melbourne}
  \country{Australia}
}

\author{Neelofar}
\email{neelofar.neelofar@monash.edu}
\orcid{0000-0003-2572-0250}
\affiliation{%
\institution{Faculty of Information Technology,}
  \institution{Monash University}
  \city{Melbourne}
  \country{Australia}
}

\author{Aldeida Aleti}
\email{Aldeida.Aleti@monash.edu}
\orcid{0000-0002-1716-690X}
\affiliation{%
  \institution{Faculty of Information Technology,}
  \institution{Monash University}
  \city{Melbourne}
  \country{Australia}
}

\date{November 2023}

\begin{abstract}
    Autonomous Vehicles (AVs) are prone to revolutionise the transportation industry. However, they must be thoroughly tested to avoid safety violations. Simulation testing plays a crucial role in finding safety violations of Automated Driving Systems (ADSs). This paper proposes PAFOT, a position-based approach testing framework, which generates adversarial driving scenarios to expose safety violations of ADSs. We introduce a 9-position grid which is virtually drawn around the Ego Vehicle (EV) and modify the driving behaviours of Non-Playable Characters (NPCs) to move within this grid. PAFOT utilises a single-objective genetic algorithm to search for adversarial test scenarios. We demonstrate PAFOT on a well-known high-fidelity simulator, CARLA. The experimental results show that PAFOT can effectively generate safety-critical scenarios to crash ADSs and is able to find collisions in a short simulation time. Furthermore, it outperforms other search-based testing techniques by finding more safety-critical scenarios under the same driving conditions within less effective simulation time.
\end{abstract}

\begin{document}
\maketitle

\section{Introduction}
In a world where the promise of self-driving cars is becoming an increasingly tangible reality, the need for thorough and reliable testing methodologies has never been more crucial. Autonomous Vehicles (AVs) must undergo rigorous testing before being deployed into the real world~\cite{barr2014oracle}. However, testing AVs directly on roads is an ineffective method, as it would require hundreds of millions of miles of test-driving to ensure their reliability~\cite{kalra2016driving}.  Furthermore, on-road testing cannot be performed for safety-critical driving scenarios.

An alternative to on-road testing is simulated testing using the agent-environment framework~\cite{feng2021intelligent}. The main idea behind simulated testing is to test the AV driving agent in a \textit{real-like} driving environment and observe its performance. Most state-of-the-art simulated testing methods leverage the naturalistic driving environment (NDE) to simulate life-like driving scenarios using high-fidelity simulators like CARLA~\cite{dosovitskiy2017carla}, BeamNG~\cite{BeamNG}, SVL~\cite{ebadi2021efficient}, NVIDIA’s Drive Constellation~\cite{nvidia2017nvidia}, and Google/Waymo’s Car-Craft~\cite{madrigal2019inside}, etc.

Given the high dimensionality of the driving environment and the rarity of the safety-critical events, exhaustive testing is infeasible and inefficient even using simulation. Therefore, the latest trend in AV testing is to employ search-based techniques that optimise the search process towards safety-critical scenarios~\cite{ben2016testing, ajdari2014adaptive, tuncali2018sim}. However, these approaches can only be applied to scenarios that involve simple manoeuvres of a limited number of vehicles. To overcome this limitation, recent studies propose modeling the driving search space in terms of a series of driving operations that other vehicles perform~\cite{li2020avfuzzer, tian2022mosat, feng2021intelligent}. In the context of AV testing, the vehicle connected to the ADS under test is referred to as the \textit{Ego Vehicle} (EV), while all other traffic participants are referred to as \textit{Non-Playable Characters} (NPCs). Considering that most safety-critical scenarios where other traffic actors are involved occur when they are located in close proximity to the EV, these techniques expend a significant portion of the optimisation process finding the driving patterns that will bring the traffic actors closer to the EV. 

This paper proposes PAFOT, a novel position-based approach for generating test cases for autonomous vehicles. We model the location of NPCs in a 9-position grid around the EV and change the driving actions of NPCs to move them among the positions. PAFOT utilises genetic algorithms to optimise these driving operations that will result in an unsafe scenario or in a collision. 

We demonstrate PAFOT using CARLA~\cite{Dosovitskiy17}, a high-fidelity simulator widely used to develop and test ADSs \cite{ding2023survey}. The experimental results show that PAFOT can effectively generate safety-critical scenarios that identify the bugs in ADSs in the form of crashes. Furthermore, when compared to state-of-the-art testing techniques, PAFOT outperforms them in terms of the number of bugs detected as well as the execution time.

To summarise, this paper makes the following contributions:

\begin{itemize}
    \item We frame our approach to testing AVs as an optimisation problem, which takes into account perturbations to the AV and other well-established metrics, such as ETTC (Estimated Time to Collision), to guide the search process. As a result, our method is capable of generating adversarial and diverse scenarios.  
    \item In order to expose safety violations of the ADS effectively during the search, we introduce a position-based approach for modifying the driving manoeuvers of NPCs, relative to the position of the EV, which increases the chance of disturbing the ADS in the generated scenarios. 
    \item We implement PAFOT and demonstrate its capabilities on CARLA, a high-fidelity simulator widely used in testing ADS. The evaluation shows that PAFOT can effectively find more safety-critical scenarios than existing techniques in less effective simulation time.
\end{itemize}

\section{Approach Overview}
\label{section:approach_overview}
\begin{figure}
    \centering
    \includegraphics[width=0.45\textwidth]{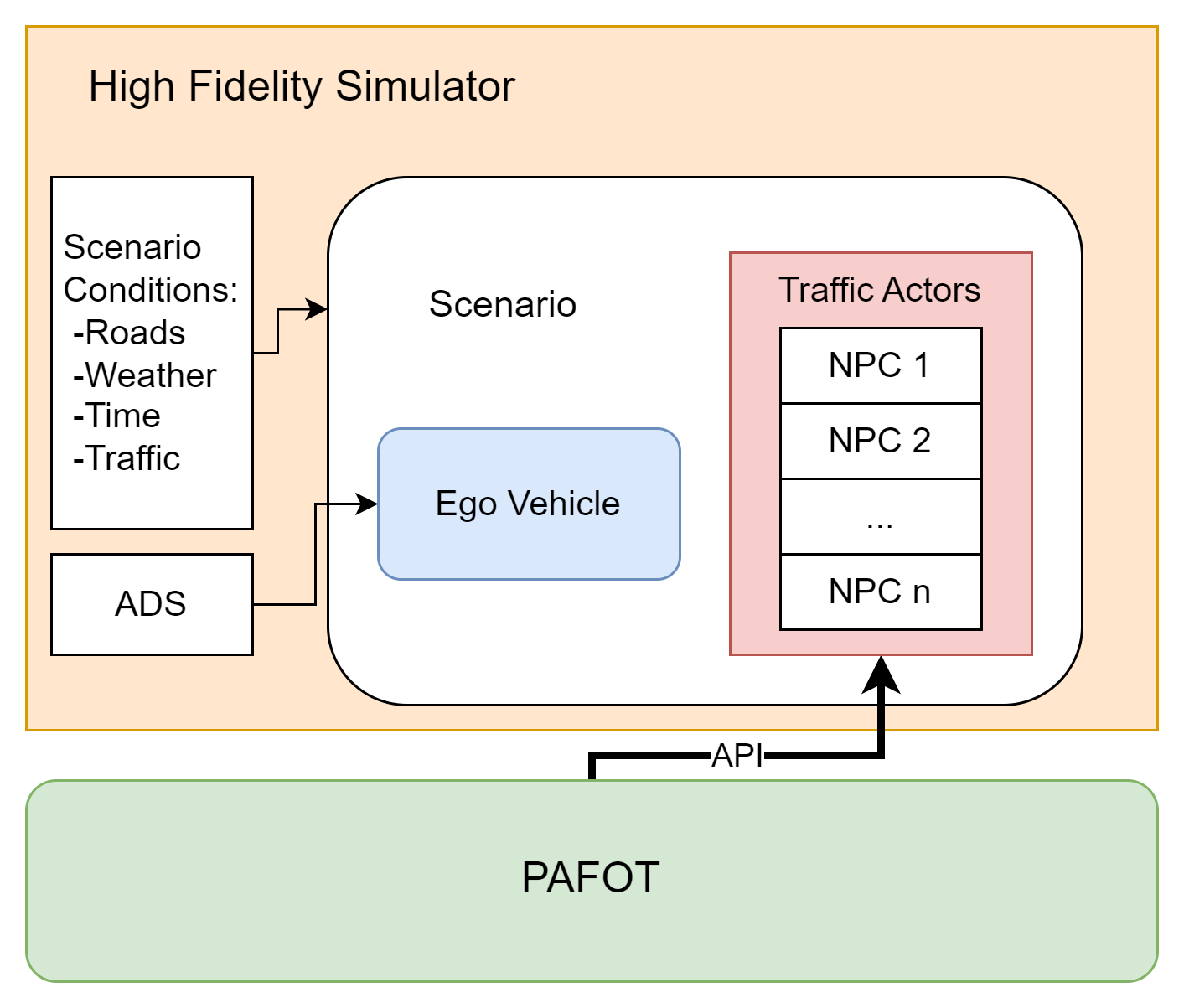}
    \caption{Overall PAFOT Framework}
    \label{fig:pafot-ads-framework}
\end{figure}

This section provides a comprehensive overview of PAFOT, describing its core objectives and offering insights into its operational framework. Its primary objective is the efficient generation of safety-critical scenarios, strategically designed to pinpoint potential safety breaches within ADS. The overall framework of PAFOT is depicted in Figure~\ref{fig:pafot-ads-framework}. PAFOT communicates with a high-fidelity simulator via an API or other means of connection that the simulator offers.

PAFOT models the NPC driving actions as high-level instructions given to the NPCs, associated with their position relative to the EV. These actions contain two main components: the target position that the NPC should move to, and the speed at which it should effectuate the action. The target position for an NPC is defined by the location of the target position relative to the NPC, as shown in Figure~\ref{fig:eight_positions}; for example, an NPC located directly in front of EV is located in target position 2, while an NPC located to the left of EV is located in target position 8. The target speed is the speed that the NPC should change to while moving to the target position. We designate the tuple of target position and speed as \textbf{Position-Instruction} (PI). Unless there is a further change of speed and/or target position, the NPC will keep its position and speed.  A scenario is composed of a set of PIs, i.e., a sequence of target positions and speeds for the NPCs to execute.

Our approach utilises a single-objective Genetic Algorithm (GA) to systematically guide the search for driving manoeuvres and patterns with the potential to compromise the safety of driving scenarios. The fitness function of such GA is primarily designed to evaluate the level of risk posed by a given scenario on the EV. Additionally, it incorporates the secondary criterion of execution time, recognising its significance in the efficiency of the proposed approach. A detailed explanation of the fitness function and the genetic operations is presented in section~\ref{sec:genetic_algorithm}.

\subsection{Modelling Target Positions}
\begin{figure}
    \centering
    \includegraphics[width=0.4\textwidth]{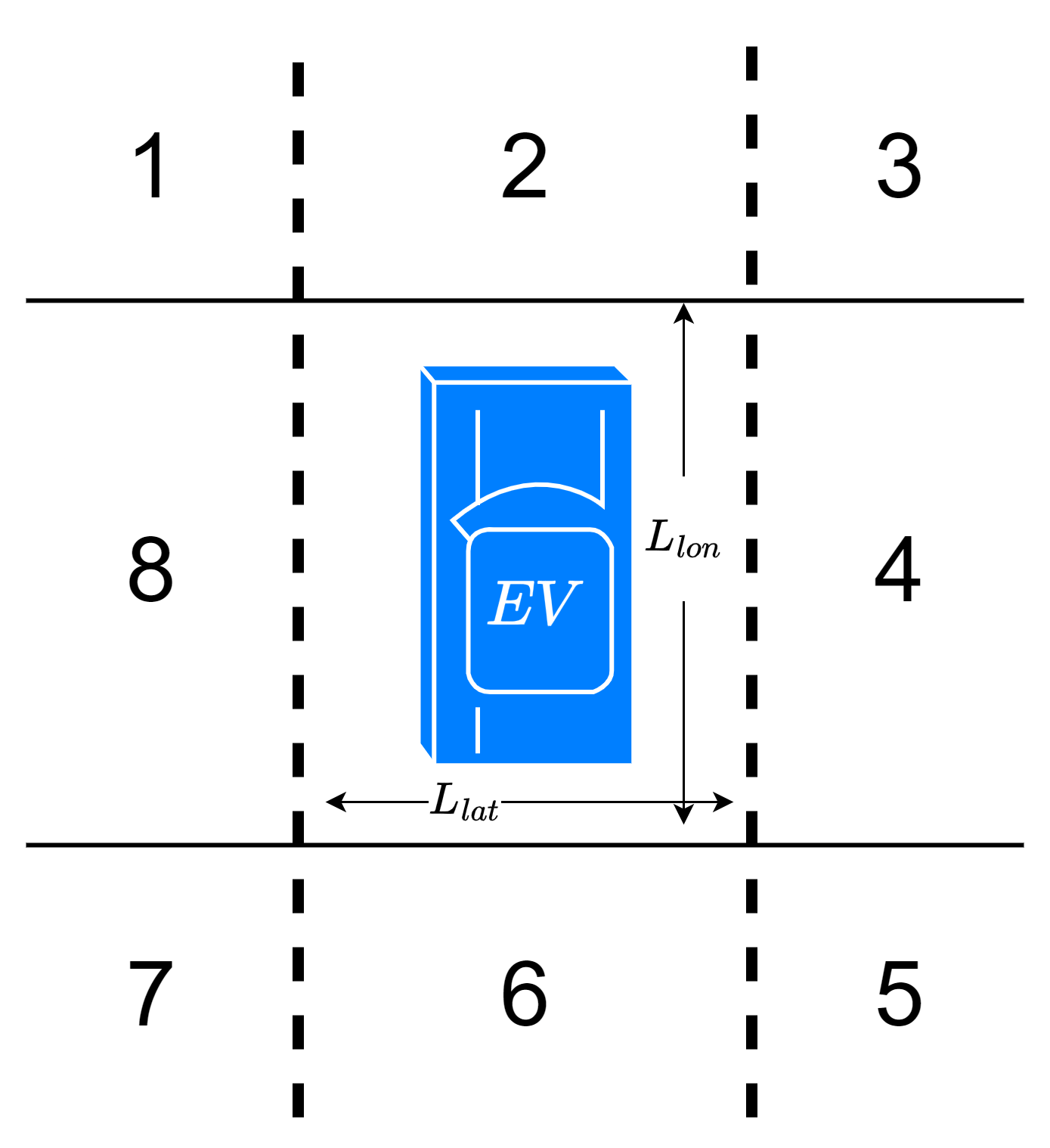}
    \caption{9-Positions grid drawn with the Ego Vehicle in the center}
    \label{fig:eight_positions}
\end{figure}
We represent the $t^{th}$ position of the $i^{th}$ NPC in the scenario as $P^i_t$. The driving actions of the NPC vehicle are determined by a sequence of these positions, denoted as $\phi^i = {P^i_1, P^i_2, ... , P^i_m}$, where m is the total number of positions in the NPC vehicle's driving pattern.

PAFOT draws a virtual 9-position grid with the Ego Vehicle in the center. The remaining eight positions are defined as target positions, identified by numbers $1$ to $8$. The dimensions of each square-position are defined by the dimensions of the Ego Vehicle's longitudinal length ( $L_{lon}$ ) and the width of the driving lane where the Ego Vehicle is located ($L_{lat}$). Each position-square has dimensions $L_{lon} \times L_{lat}$. The geometrical center of each square-position is where the target position is located. The type of operations required to situate the NPC in the target position are not supported out-of-the-box by the CARLA Simulator. Therefore, we implemented a subroutine, based on a PID controller~\cite{PIDcontroller}, to move the NPC to a desired waypoint located in the geometrical center of the target position. Figure~\ref{fig:eight_positions} presents the graphical representation of the 9-positions grid virtually drawn, and the identifier for each square-position.

\begin{figure}[H]
    \centering
    \includegraphics[width=0.5\textwidth]{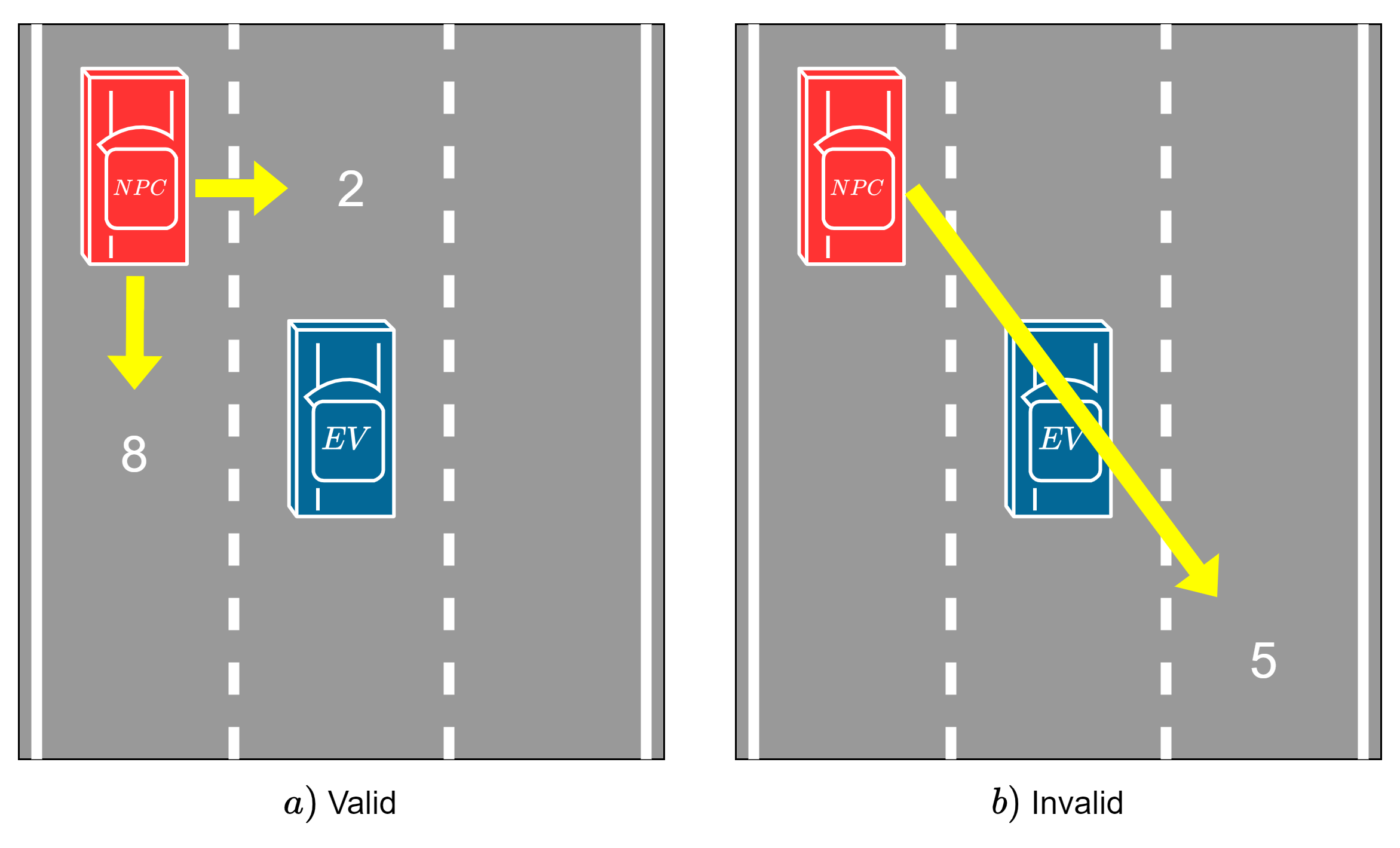}
    \caption{Valid and invalid movements for an NPC situated in Position 1}
    \label{fig:valid_invalid_movements}
\end{figure}

Note that, because of the way PAFOT models the position of the NPCs relative to the EV, there might exist sequences of PIs that are either physically impossible, invalid, or non-naturalistic. Figure~\ref{fig:valid_invalid_movements}-b shows an invalid sequence of PIs for an NPC situated in Position 1 (i.e., from Position 1 move to Position 5). PAFOT introduces a mechanism to ensure that the sequences of PIs are physically possible (i.e., will not \textit{teleport} the NPCs between square-positions) and naturalistic (i.e., will not make the NPCs perform actions such as moving sideways, etc). At every step of the simulation, PAFOT checks whether the next Position is adjacent to the current; in case the next Position is not adjacent, PAFOT will select an adjacent Position with a probability of 50\% for each adjacent Position, and update the solution to reflect this change. This mechanism is also employed for occasions where a target position is not available (e.g., there is not a road lane on either side of the Ego Vehicle, etc). Figure~\ref{fig:valid_invalid_movements}-a shows the valid sequence of PIs for an NPC situated in Position 1 (i.e., from Position 1 move to either Position 2 or Position 8). 

\section{PAFOT}
\label{sec:scenario_representation}
In this section, we introduce our approach to modelling the generation of safety-critical scenarios as an optimisation problem and describe into the genetic representation of test scenarios. To ensure the comprehensiveness of this section, we provide a brief explanation of the Genetic Algorithm and its components at the outset.

\subsection{Genetic Algorithm}
\label{sec:genetic_algorithm}

Genetic Algorithms (GAs) are meta-heuristic search algorithms inspired by natural evolution. The search starts with an initial set of candidate solutions (i.e., testing scenarios) that are collectively called a \textit{population}. The search is guided by a \textit{fitness function}, which calculates the fitness score of each candidate. This fitness score represents the performance of a candidate at solving the problem.

At each iteration, some candidate solutions are selected from the population for recombination operations, involving two types: (a) crossover and (b) mutation. In crossover, two candidates are randomly chosen and swapped, aiming to find an improved solution from an already favourable one. This operation typically refines the search, moving closer towards an optimal solution. Mutation randomly selects one candidate and mutates or changes part of the solution, thereby broadening the algorithm's search exploration. Recombination operations, in general, give rise to novel, higher-performing members, which are then integrated into the population. On the other hand, members with lower fitness scores are gradually phased out. Each iteration is generally called a \textit{generation}, and this process repeats until either a population member achieves a desired fitness score (indicating a solution was found) or the algorithm concludes after exceeding the allocated time limit.

\begin{figure}
    \centering
    \includegraphics[width=0.5\textwidth]{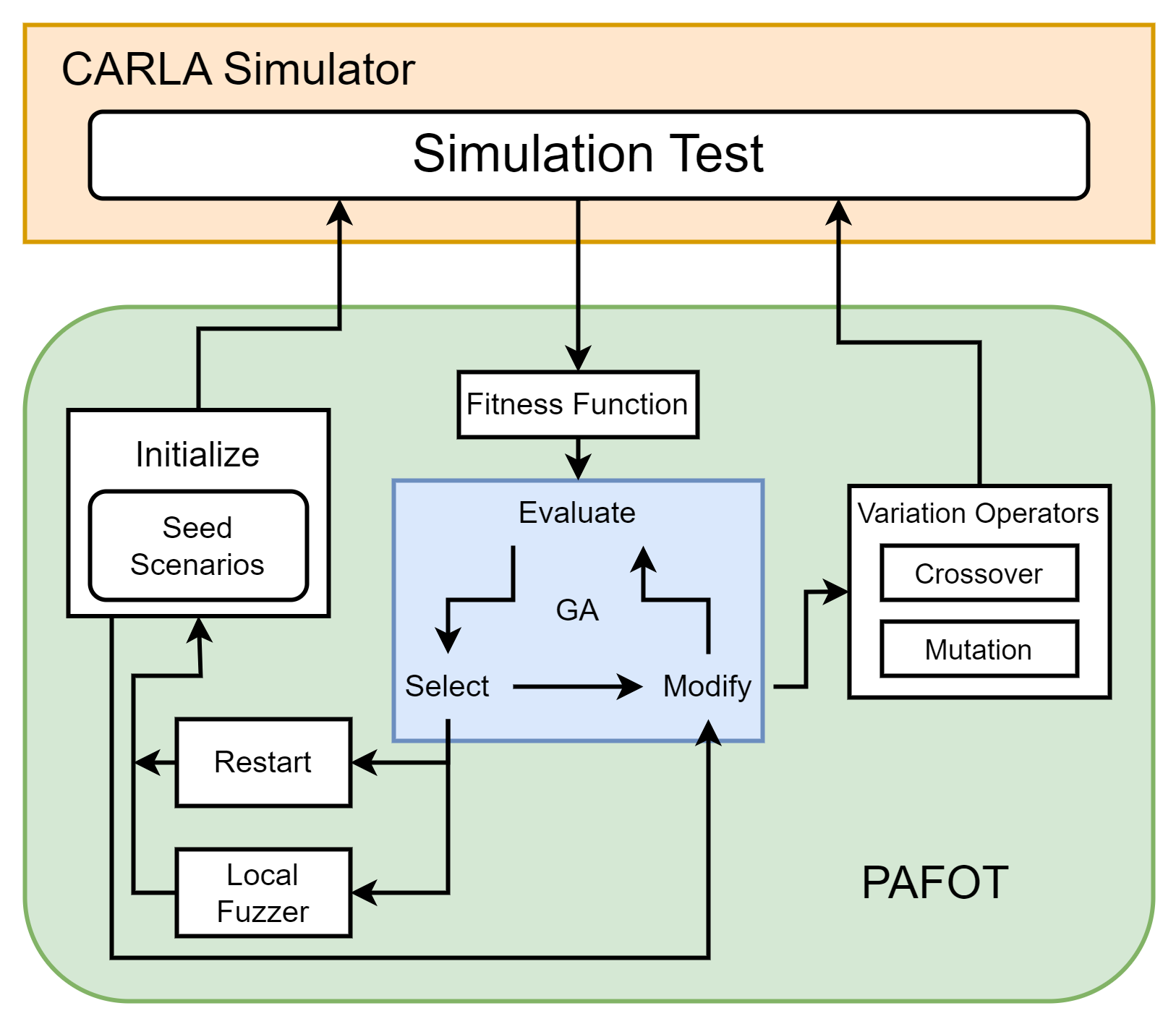}
    \caption{Genetic representation of PAFOT}
    \label{fig:pafot_framework}
\end{figure}

The high-level workflow of the genetic algorithm in PAFOT is shown in Figure~\ref{fig:pafot_framework}.

\subsection{Genetic Representation of a Test Scenario} 
\label{sec:genetic_scenarios}
\begin{figure}
    \centering
    \includegraphics[width=0.5\textwidth]{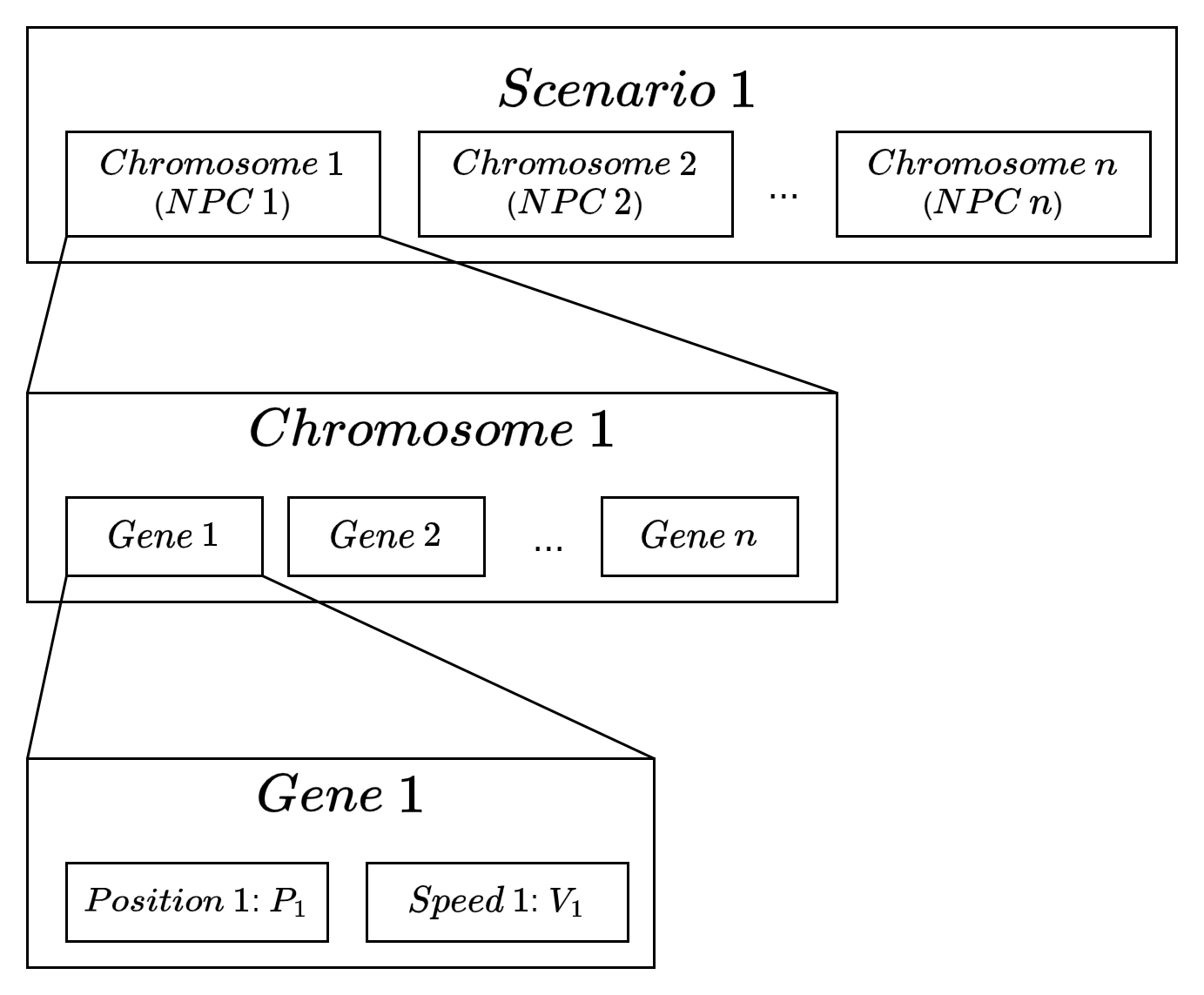}
    \caption{Genetic representation of a testing scenario}
    \label{fig:scenario_representation}
\end{figure}

PAFOT disrupts the driving manoeuvres and patterns of NPCs in a continuously evolving environment to introduce challenges for the EV in a test scenario. A test scenario is represented as an individual within a genetic algorithm.  Figure~\ref{fig:scenario_representation} presents a visual depiction of a test scenario.

Each individual or test scenario consists of one or more chromosomes  $C_i = \{C_1, C_2, ..., C_n\}$. Each chromosome represents an NPC in a scenario. PIs are encoded as the genes of the chromosome, conformed by a target position ($P_j$) and a target speed ($V_j$). When a new evolution starts, PAFOT randomly chooses \textit{k} individuals, consisting of a sequence of valid PIs, as the initial population.

\subsection{Fitness Function}
The fitness function of the GA is used to determine the individuals that are to be carried over for the next evolution. The fitness function employed by PAFOT gives each individual a fitness value calculated on the basis of four performance metrics: the minimum estimated time to collision (METTC), the minimum distance (MD) between the EV and an NPC, the safety distance (SD) between the EV and surrounding NPCs, and the total execution time of an individual scenario. PAFOT minimises all these values to find optimal tests. We perform a linear combination of these values to find the fitness score of an individual.

\begin{itemize}
    \item Minimum Estimated Time to Collision (METTC): refers to the smallest anticipated duration until a collision is expected to occur between EV and the NPCs in the test scenario~\cite{schwarz2014computing} if both keep the same speed and trajectories over time~\cite{van1994time}. Given that $(x_1, y_1)$ is the position of the EV, ($x_2, y_2$) is the position of the NPC, $(x_+, y_+)$ is the estimated collision point at the time-stamp $t$, $V_1$ is the speed of the EV, and $\theta_1$ and $\theta_2$ are the angles between the direction of the driving of the EV and NPC, and the road direction, Estimated Time to Collision (ETTC) is calculated as:
    \vspace{3pt}
    \begin{equation}
        ETTC = \frac{\sqrt{{(y_+ - y_1)}^2 + {(x_+ - x_1)}^2}}{|V_1|}
        \end{equation}
        \vspace{3pt}
        while $x_+$ and $y_+$ are calculated as:
        \vspace{3pt}
    \begin{equation}   
        x_+ = \frac{y_2 - y_1 + x_1tan\theta_1 - x_2tan\theta_2}{tan\theta_1 - tan\theta_2}
    \end{equation}
    \vspace{3pt}
    \begin{equation}
        y_+ = \frac{x_2-x_1 + y_1cot\theta_1 - y_2cot\theta_2}{cot\theta_1 - cot\theta_2}
    \end{equation}
    \vspace{3pt}
    \item Minimum distance (MD): is defined as the minimum Euclidean distance between the EV and an NPC, and calculated with Equation~\ref{eq:distance}, where ($x_1, y_1$) and ($x_2, y_2$) are the Cartesian coordinates of the EV and NPC respectively. The smaller this distance, the higher will be the fitness score.
    
    \begin{equation}
    \label{eq:distance}
        D = \sqrt{{(x_2 - x_1)}^2  + {(y_2 - y_1)}^2}
    \end{equation}
    \vspace{2pt}
    \item Safety Distance (SD): is defined as the distance that a vehicle must maintain with another vehicle in its close vicinity to ensure there is enough time to manoeuvre should an unsafe driving scenario occur. Based on the literature, the recommended time to manoeuvre is between 2 and 6 seconds~\cite{curry2021normaldriving, wu2017minimumsafetydistance}. Therefore, we adopt a time of 3 seconds as SD in our study. Like other distances, this distance needed to be minimised for a scenario to be safety-critical.

    \begin{equation}
        SD = (v_1 - v_2) t + \frac{1}{2}(a_1-a_2)t^2
    \end{equation}

    where $v_i$ and $a_i$ are the speed and acceleration of the vehicles.
    
    \item Execution Time (ET): is defined as the time it takes for a scenario to find a safety violation. The smaller the execution time, the higher the fitness score.
    \begin{equation}
        ET = t_f - t_0
    \end{equation}
\end{itemize}

In our simulations experiments, we measured a provisional fitness function of an individual test scenario six times per second, and the final fitness value is calculated once the scenario is finalised to include $ET$ into the calculations. The fitness score is used to rank each test scenario, and the GA decides whether the candidate should be kept for the next round of evolution or discarded.

\subsection{Variation Operators}
The variation operators consist of two types: mutation and crossover. 

\subsubsection{Crossover}

\begin{figure}[H]
    \centering
    \includegraphics[width=0.45\textwidth]{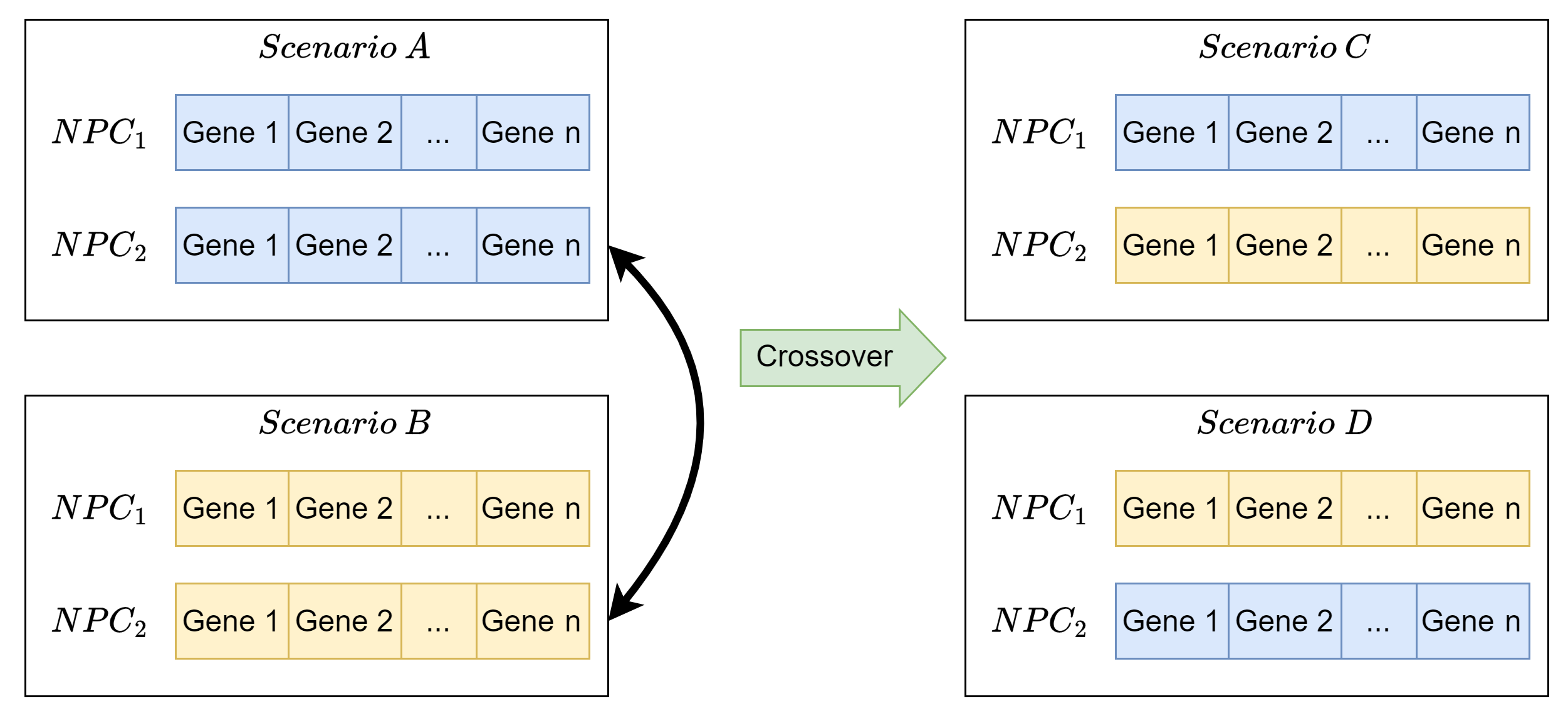}
    \caption{Crossover of two scenarios}
    \label{fig:crossover}
\end{figure}

The crossover operation consists of a swap operation. It aims to increase the probability of combining NPCs from two scenarios to form a new scenario with a higher chance of finding a safety violation. In each generation, the swap operation randomly chooses two NPCs, one from each scenario, and swaps the two NPCs, with a probability of $p_c$. Figure~\ref{fig:crossover} shows an example of a crossover operation between two individuals to form two new scenarios. This process contributes to the genetic diversity within the population, inducing the exploration of new combinations of driving behaviours and enhancing the overall potential of the genetic algorithm in uncovering safety-critical scenarios.

\subsubsection{Mutation}

\begin{figure}[H]
    \centering
    \includegraphics[width=0.45\textwidth]{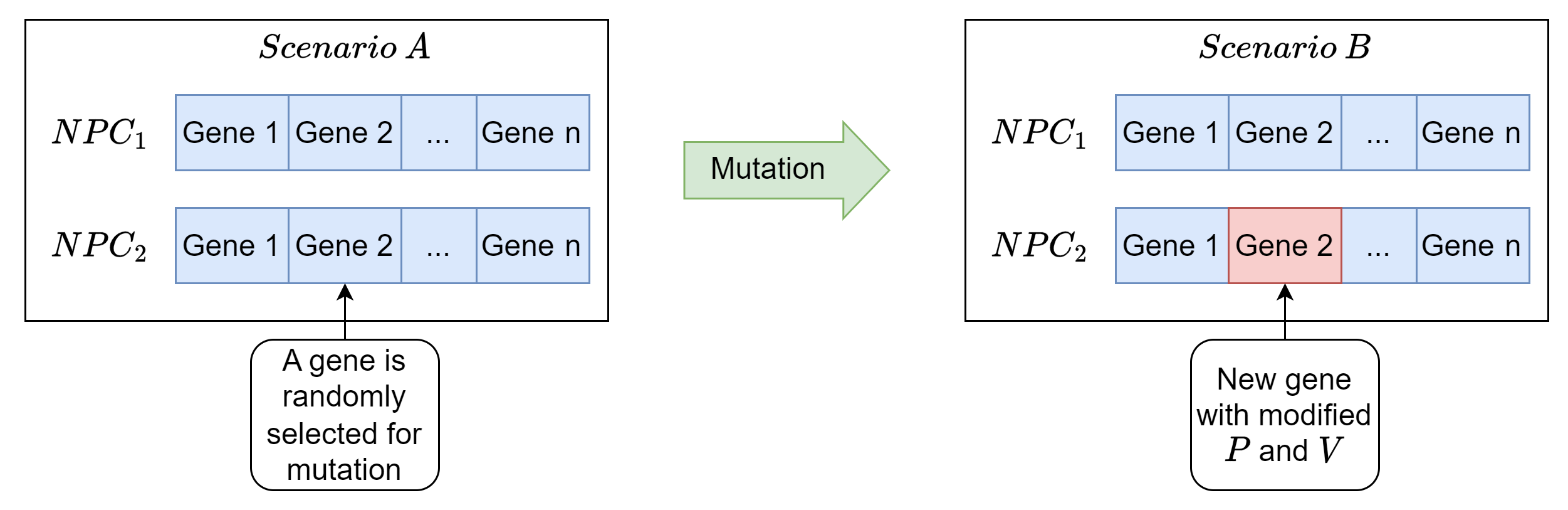}
    \caption{Mutation operation in one scenario}
    \label{fig:mutation}
\end{figure}

Given a set of PIs as genes in a chromosome, the mutation operation selects one of them at random and changes it with a new PI with a probability of $p_m$. PAFOT then engages the mechanism presented in section~\ref{section:approach_overview} to ensure that all PIs are valid. Figure~\ref{fig:mutation} presents an example of a mutation operation performed on an individual to form a new scenario.

We experimented with $p_m$ and $p_c$ values in the range that the literature recommends~\cite{haupt2000optimum} and chose 0.5 for both of them. These values resulted in the shortest time to encounter safety violations and collisions in our simulations.

\subsection{Selection}
In our genetic algorithm implementation, we chose to employ an \textit{Elitist} strategy~\cite{purshouse2002use} for the selection of individuals to advance to the subsequent generation. The Elitist strategy involves keeping a small proportion of the top-performing individuals from the previous generation, thereby preserving and reinforcing valuable traits. In our specific implementation, we opted to incorporate a measured form of elitism by selectively carrying over the top 25\% of the elite individuals to the next generation. By integrating this elitist selection approach, our genetic algorithm attempts to maintain a constructive balance between continuity and innovation in successive generations, optimizing the overall evolutionary process towards the generation of safety-critical scenarios.

\subsection{Local Fuzzer}

The purpose of the local fuzzer is to dynamically increase the exploration of the surrounding search areas when PAFOT identifies scenarios with a significant likelihood of safety violations. These high-potential cases might indicate near-miss cases of safety breaches, and we utilise them as initial seed scenarios.

The local fuzzer process focuses on exploiting the local optima within the neighbourhood, using the seed scenario collected by the GA. When the GA identifies a scenario with a high fitness score after a set of stages of evolution, a sub-routine of the local fuzzer is triggered and the main GA's process is paused. In the local fuzzer process, a fresh population is initialised with the seed scenario previously identified, and a mutation-based fuzzing process is carried out using this population. To maintain a significant potential for safety violations during the local fuzzing, we adhere to the same fitness function as used in the GA, but with an increased mutation rate. Upon completing the local fuzzing, if a scenario with a higher fitness score is discovered, it replaces the seed scenario in the GA, allowing for the introduction of a promising scenario into the GA population. After the local fuzzing is completed, the sub-routine concludes, and the primary GA process resumes.

\subsection{Restart}
\label{subsec:restart}

To supplement the capabilities of the local fuzzing engine, PAFOT incorporates a random restart mechanism designed to enhance solution. This mechanism comes into action as a reactive measure, moving the search away from areas of the search space that may have become stagnant or thoroughly explored by the GA. The random restart is activated when a lack of improvement in fitness score is observed over a defined period of time or across multiple generations. This process serves as a dynamic strategy to redirect the search to unexplored regions within the search space by introducing an element of randomness. By dynamically enforcing random restarts, PAFOT not only counteracts potential convergence to suboptimal solutions but also maintains a balanced exploration-exploitation trade-off, ensuring a more comprehensive and effective search for safety-critical scenarios.

\subsection{Termination}

The iterative process of the GA, local fuzzer, and random restart is repeated until a given number of generations is fully explored. After the completion, PAFOT compiles and returns a list of scenarios wherein safety violations were observed, denoted by the occurrence of collisions during the process.

\section{Experimental Design and Results}

In this section, we execute PAFOT to generate safety-critical scenarios for testing the CARLA Simulator's built-in ADS. To evaluate the efficiency and effectiveness of PAFOT, we seek to answer the following research questions:

\begin{itemize}
    \item \textbf{RQ1}: How effective is PAFOT at generating test cases that expose safety violations of ADSs?
    \item \textbf{RQ2}: How efficient is PAFOT when compared to existing techniques for testing ADSs?
\end{itemize}

\subsection{Experiment Setup}

We conducted our experiments on Ubuntu 18.04.6 LTS with an AMD Ryzen 5 5600G, 32GB of RAM Memory, and an NVIDIA RTX 3070. The CARLA Simulator~\cite{Dosovitskiy17} was selected as the simulation platform for conducting our experiments, using its built-in traffic manager ADS for controlling the EV. We \textit{spawn} three actors, one of which is treated as the EV and has been attached with the CARLA's traffic manager ADS for controlling it, while the remaining two are the NPCs and controlled by PAFOT. 

\begin{figure}[H]
    \centering
    \includegraphics[width= 0.45\textwidth]{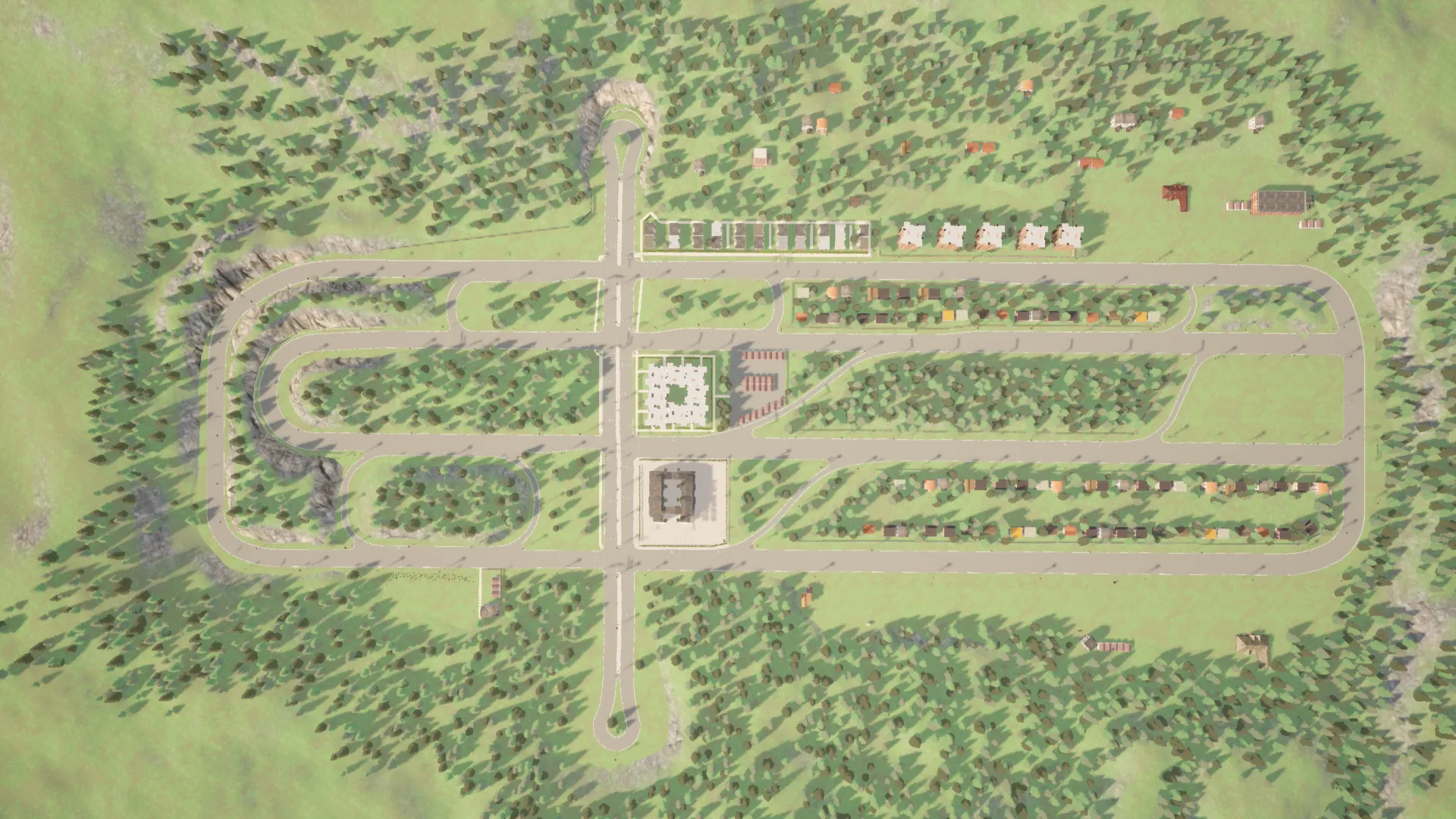}
    \caption{Aerial view of Town 06 Map}
    \label{fig:map-aerial}
\end{figure}

We selected the Town 06 Map from the list of available maps on CARLA, as it provides a diverse driving environment including a mix of highway and urban road conditions, containing intersections, merging and diverging lanes, and Michigan Left. The map Town 06 is shown in Figure~\ref{fig:map-aerial}. While we used the Town 06 map for our experiments, PAFOT is map-independent and can be implemented with other maps and/or high-fidelity simulators. 

The operational design domain in our experiments includes the following conditions:
\begin{itemize}
    \item Weather: Sunny during daytime
    \item Traffic and road components: Traffic lights, lane markers on roads, pedestrian crossings. The road ranges from 4 to 6 lanes.
    \item Speed limits: The speed limit of the NPCs and the EV is 60 mph ($\sim$ 96 km/h). 
    \item Number of NPCs: We set each testing technique with a fixed number of 2 NPCs. We found that, given the nature of our approach, 2 NPCs cause sufficient traffic on the road without crowding the close vicinity of the EV.
\end{itemize}

PAFOT is primarily focused towards addressing a key safety requirement in the testing of ADSs; it being the prevention of collisions with other participants in different scenarios. The focal point of this approach is calculating METTC to assess the collision risk of the EV. Beyond collision avoidance, PAFOT incorporates a spectrum of additional safety considerations, including adhering to traffic rules. This comprises safely navigating scenarios, such as avoiding pedestrians, adhering to traffic lights, and accurately following lane directions, among other regulatory traffic rules.

\subsection{Experiment Design}
We run PAFOT to generate scenarios and evaluate the performance of the ADS in the test scenarios. To monitor the occurrences of collisions between the EV and other actors or objects in the simulated environment, a collision sensor is attached to the EV. This sensor sends a signal to PAFOT which stops the simulation and records the scenario.

Following its implementation, PAFOT undergoes a comparative analysis against two established baselines: AV-Fuzzer~\cite{li2020avfuzzer} and Random. AV-Fuzzer, a noteworthy open-source search-based testing technique, employs genetic algorithms to systematically evolve the manoeuvres of NPCs with the explicit goal of uncovering safety violations of ADSs. The inclusion of AV-Fuzzer as a baseline provides a benchmark for assessing the effectiveness and efficiency of PAFOT in generating safety-critical scenarios. Alongside PAFOT and AV-Fuzzer, a Random testing technique is introduced as an additional baseline. This method introduces a controlled level of randomness by randomly manipulating the driving behaviours of NPCs at specified intervals during the simulation. By comparing PAFOT against both AV-Fuzzer and Random, our evaluation aims to outline the unique contributions and advantages offered by PAFOT, demonstrating its efficacy and efficiency in generating safety-critical scenarios for the evaluation of ADSs.

We run these three testing techniques for 10 times, and compare the effectiveness and efficiency in the following specifics:
\begin{itemize}
    \item How many test scenarios are generated?
    \item How many collisions are found in each run of the techniques?
    \item How long does it take to find a collision?
    \item How much time does it take to run all the experiments?
\end{itemize}

\subsection{Result Analysis: RQ1}

We run PAFOT, AV-Fuzzer, and Random on the same road location, which includes highway and urban driving conditions. The experiments were repeated 10 times, to account for the stochastic nature of some components of the testing techniques. Each experiment was given 100 generations to perform the evolution or, in the case of the Random, was repeated 100 times with a randomly generated set of scenarios. The iterations in the experiments were crucial for ensuring a comprehensive assessment of the testing techniques across multiple driving conditions, and to account for the inherent randomness in certain components of the techniques, and the robustness of the methodologies over multiple repetitions. 

\begin{figure}[H]
    \centering
    \includegraphics[width=0.45\textwidth]{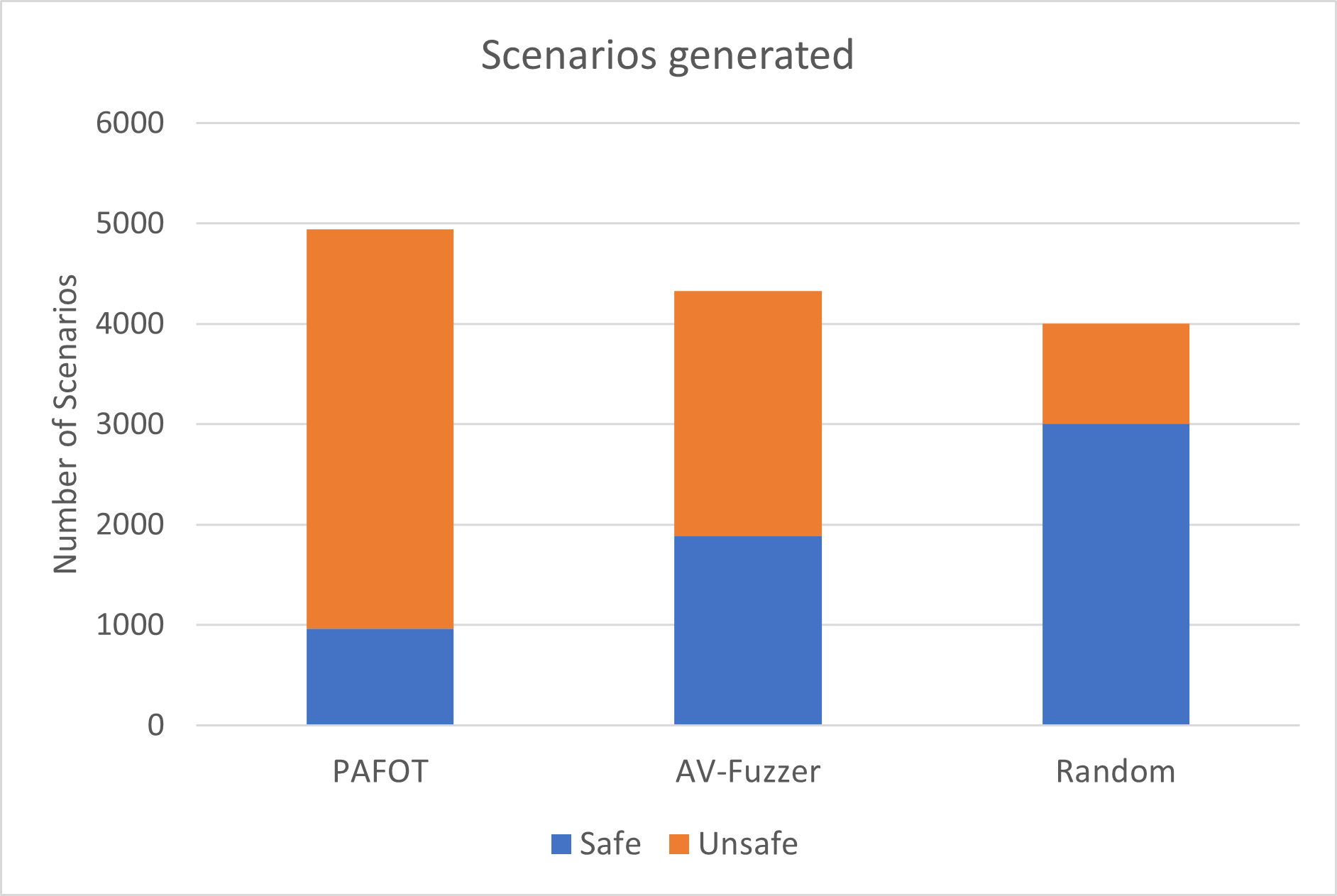}
    \caption{Total scenarios generated by each technique}
    \label{fig:total_scenarios}
\end{figure}

PAFOT was able to generate and run 4945 test scenarios in total, of which, 3981 have safety violations. AV-Fuzzer generated 4327 scenarios in total, of which 2438 resulted in a collision. Random generated 4000 scenarios, of which 997 resulted in a collision. Figure~\ref{fig:total_scenarios} shows the total scenarios and their classification that were generated in total by each testing technique. It is evident that the number of scenarios generated is bigger on PAFOT than on AV-Fuzzer and Random Fuzzer. 

The difference in total scenarios generated can be explained by the different number of times that each testing technique engaged the Local Fuzzer. Remember that, as explained in section~\ref{sec:genetic_algorithm}, the objective of the Local Fuzzer is to increase the variation rates to find more scenarios with collision potential around an area of the search space where the testing technique finds safety violations. As a consequence of performing a local search, more test scenarios are generated. PAFOT engaged the Local Fuzzer 185 times in total, while AV Fuzzer performed 80 of these local searches. The Random Fuzzer, due to the nature of the technique, did not perform local searches during its execution. Table~\ref{tab:scenarios_comparison_collisions} shows the comparison of the number of total scenarios and collisions that each technique found.

\begin{table}[H]
    \centering
    \caption{Number of scenarios generated}
    \begin{tabular}{c|c|c|c}
    \hline
    & PAFOT & AV-Fuzzer & Random\\
    \hline
     Scenarios generated & 4945 & 4327 & 4000 \\
     Collisions & 3981 & 2438 & 997 \\
     \hline
    \end{tabular}
    \label{tab:scenarios_comparison_collisions}
\end{table}

\begin{figure}[H]
    \centering
    \includegraphics[width=0.45\textwidth]{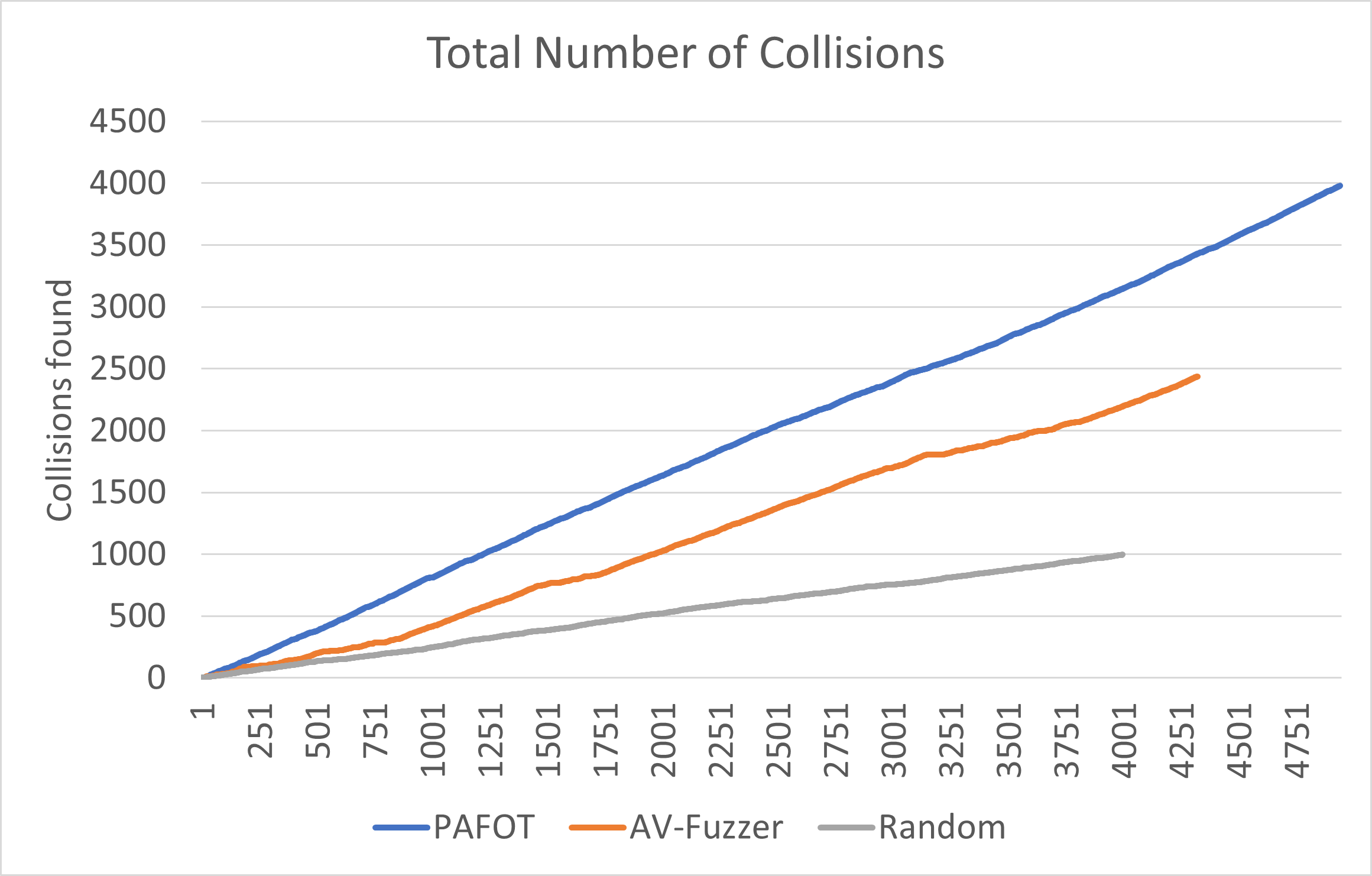}
    \caption{Number of collisions found over time by each testing technique}
    \label{fig:number_collisions}
\end{figure}

The dynamic evolution of the number of collisions detected by each testing technique is visually depicted in Figure~\ref{fig:number_collisions}. Notably, PAFOT distinguishes itself by showing a steeper trajectory in the progression of the number of collisions found over time when compared to both AV-Fuzzer and Random. While all three approaches effectively succeed in detecting collisions, their individual rates of detection exhibit substantial variations. The graph captures the nuanced temporal dynamics of collision identification, portraying how PAFOT tends to identify collisions at a more accelerated pace than the other testing techniques. This difference in the temporal patterns of detecting collisions prompts a closer examination, and a further analysis of these observations will be conducted in section \ref{subsec:rq2_analysis}.

\begin{figure}[H]
    \centering
    \includegraphics[width=0.45\textwidth]{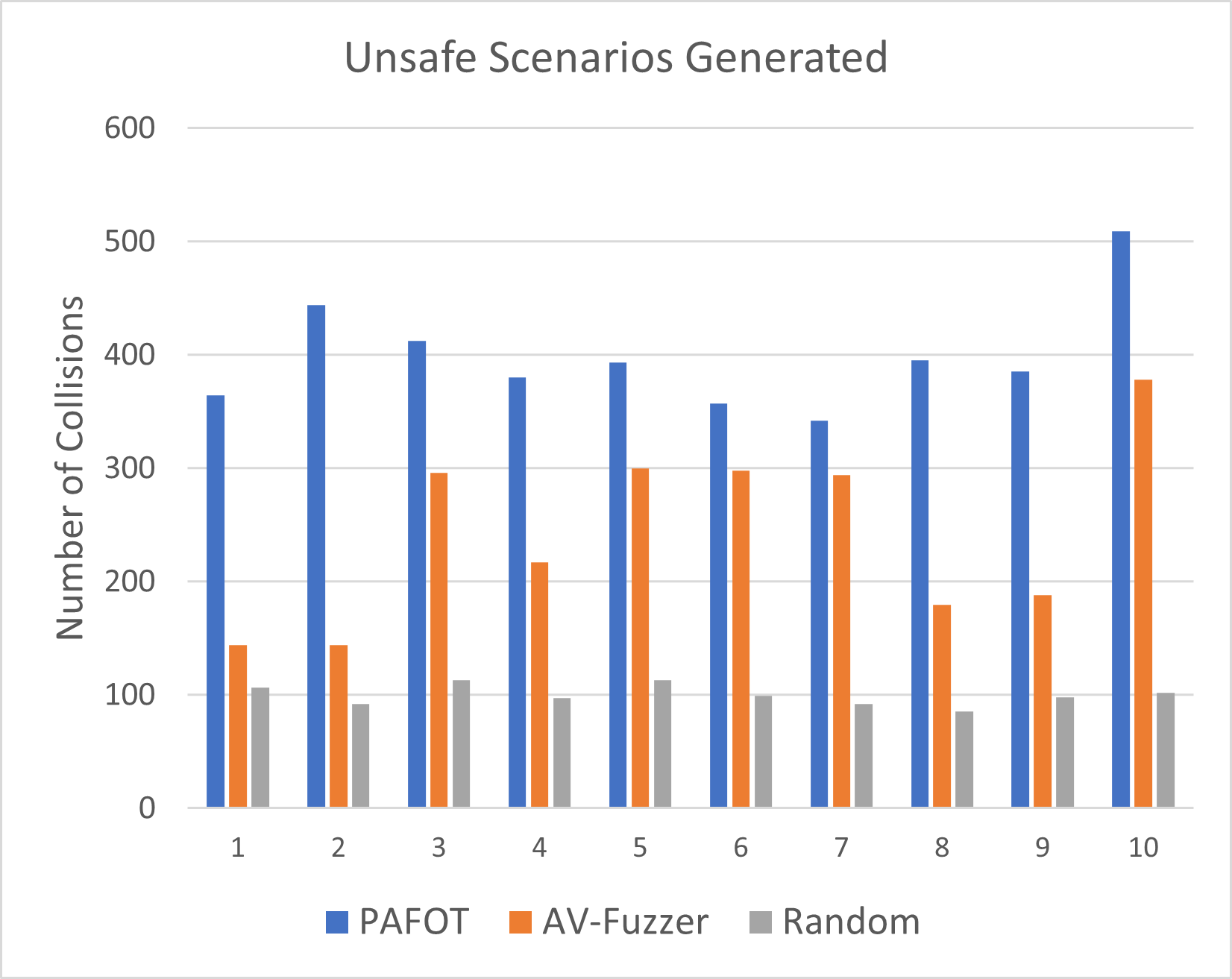}
    \caption{Unsafe scenarios generated by each technique}
    \label{fig:unsafe_scenarios_each_technique}
\end{figure}

An in-depth analysis of the results across the 10 times the experiments were executed reveals a clear and consistent trend, highlighting the capability of PAFOT to generate safety-critical scenarios across multiple repetitions. Figure~\ref{fig:unsafe_scenarios_each_technique} provides a visual breakdown of the number of both safe and unsafe scenarios generated by each testing technique in each of the 10 individual runs. Notably, Random appears as the technique with the least number of unsafe scenarios in each run. In contrast, PAFOT consistently outperforms both AV-Fuzzer and Random in generating safety-critical or unsafe scenarios. On average, PAFOT generates 398 unsafe scenarios per run, with a maximum of 509 and a minimum of 342, AV-Fuzzer generates 243 unsafe scenarios on average per run, with a maximum of 378 and minimum of 144, and Random generates 99 unsafe scenarios on average per run, with a maximum 113 and minimum 85.  

\begin{custombox}[Answer to RQ1] Across 10 runs, PAFOT successfully generated a total of 4945 test scenarios, and 3981 of these scenarios resulted in collisions, marking a significant improvement over the outcomes achieved by AV-Fuzzer and Random. Specifically, when compared to AV-Fuzzer, which generated 2438 collision scenarios, PAFOT demonstrated an improvement of 1543 safety-critical scenarios. Furthermore, in comparison to Random, which produced 997 safety-critical scenarios, PAFOT showcased an even more substantial improvement of 2984 scenarios that resulted in a collision. The experimental results show PAFOT's effectiveness in identifying safety-critical scenarios, surpassing both AV-Fuzzer and Random.
\end{custombox}

\subsection{Result Analysis: RQ2}
\label{subsec:rq2_analysis}
We are similarly interested in understanding how efficient PAFOT is, this is, the time that it takes to find a collision and its comparison with other testing techniques.

On average, the execution time for a single test scenario in PAFOT, regardless of its outcome, is 28.16. A test scenario that does not result in a collision is stopped after the allocated time of 60 seconds is exhausted. On the contrary, the average execution time of an unsafe scenario, this is, the time it took PAFOT to find a collision is 20.65 seconds. AV-Fuzzer takes 45.53 seconds to run a single test scenario, while it takes on average 34.32 seconds to find a collision in an unsafe scenario. Random, on the other hand, takes 54.42 seconds on average to execute a scenario, and 37.63 seconds to find a collision in an unsafe scenario. The difference in the time that the testing techniques require to find a collision might help to explain the rate at which each technique finds collisions over time. 
Table~\ref{tab:scenarios_comparison} shows the results of the three testing techniques, after 10 runs of the experiments.

\begin{figure}[H]
    \centering
    \includegraphics[width=0.45\textwidth]{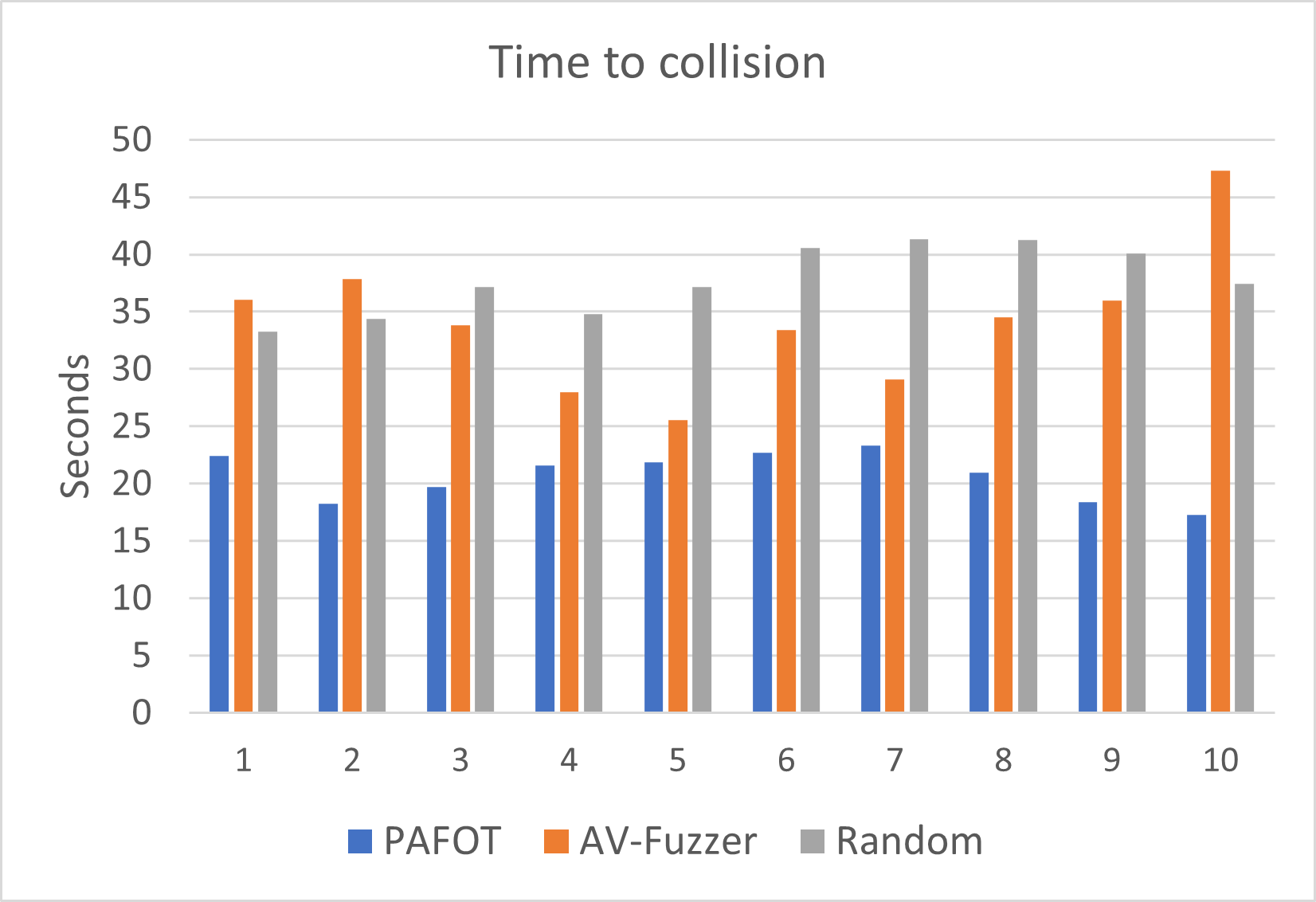}
    \caption{Time to collision for each experiment run}
    \label{fig:time_collision}
\end{figure}

The time that it takes a testing technique to find a safety violation is a metric that similarly prevails across each of the 10 runs that each technique executed. PAFOT takes 20.65 (max 60 and min 3) seconds on average to find a collision in an unsafe scenario, while AV-Fuzzer takes 34.32 seconds (max 60 and min 11), and Random Fuzzer takes 37.63 seconds (max 60 and min 10). Figure~\ref{fig:time_collision} shows how each testing technique performed on the average time to collision on each run of the experiments. It is important to remember that this time is only relevant when there is a collision found since a safe scenario without a collision will end when the total allocated time of 60 seconds per scenario is exhausted. From these results, one can conclude that PAFOT takes less time on average to find a collision in an unsafe scenario.

\begin{figure}[H]
    \centering
    \includegraphics[width=0.45\textwidth]{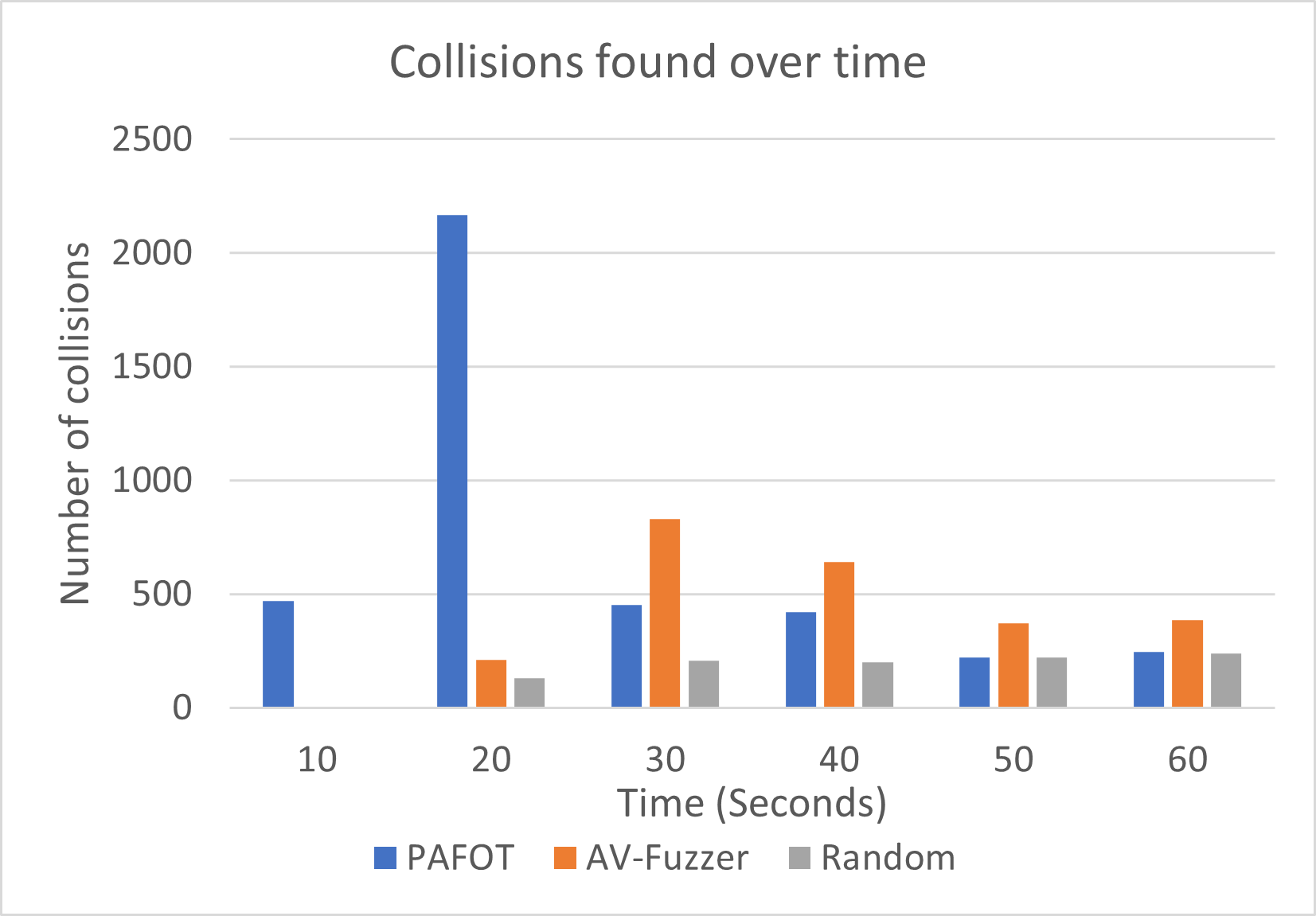}
    \caption{Time in the simulation where a collision happened}
    \label{fig:collisions_over_time_all}
\end{figure}

The timing of a collision within a scenario adds an additional layer of interest to the results. The information is visually represented in Figure~\ref{fig:collisions_over_time_all}, where the x-axis shows time intervals in the simulation, categorised into 10-second increments. PAFOT exhibits a distinctive pattern, with the majority of collisions occurring between 10 and 20 seconds after the simulation started. In contrast, AV-Fuzzer tends to identify most collisions in a slightly later phase, primarily between 20 and 30 seconds into the simulation. Notably, Random follows a different trend, with the preponderance of collisions slightly concentrated in the final 10 seconds of the simulation, specifically between 50 and 60 seconds after the start. This can offer insights as to how the position-based approach that is introduced with PAFOT tends to find more collisions within the first 20 or 30 seconds into the simulation. In addition to this, the intervals between 40 to 60 seconds are where the least amount of collisions were found by PAFOT. In a similar analysis, AV-Fuzzer is able to find most collisions in the interval between 20 and 40 seconds after the simulation commenced. Surprisingly, it was unable to find collisions within the first 10 seconds of the simulation. Random finds most collisions in the final stages of the simulation. However, collisions are uniformly distributed across all time intervals after 10 seconds.

While simulation-based testing shows an improvement in the time that it takes to test an ADS when compared to other testing approaches, it is important to remember that simulations are expensive in terms of execution time and computational resources. Therefore, it is important to find more safety-critical scenarios in less simulation time. PAFOT shows an improvement of 13.67 seconds to find a collision, compared to AV-Fuzzer, and 16.98 seconds compared to Random. While it might seem that this improvement in execution time is small, it is evident when we compare the total execution time that each testing technique took to execute 10 runs of the experiments. PAFOT took in total 38.68 hours to complete all the experiments, while AV-Fuzzer took 54.73 hours, and Random took 60.47 hours. This means, PAFOT shows an improvement of 16.05 hours in total execution time, when compared to AV-Fuzzer and 21.79 hours when compared to Random. Table \ref{tab:scenarios_comparison} shows a comparison, in terms of time, of all three testing techniques.

\begin{table}[H]
    \centering
    \caption{Comparison of testing techniques}
    \begin{tabular}{c|c|c|c}
    \hline
    & PAFOT & AV-Fuzzer & Random\\
    \hline
     Time to collision (avg) & 20.65 $s$ & 34.32 $s$ & 37.63 $s$\\
     Total execution time (hours) &38.68& 54.73 & 60.47\\
     \hline
    \end{tabular}
    \label{tab:scenarios_comparison}
\end{table}

\begin{custombox}[Answer to RQ2]
    After 10 runs, on average, PAFOT is able to generate and run one effective test case, this is, that results in a collision, 13.67 seconds faster than AV-Fuzzer and 16.98 seconds when compared to Random. When taken into account all scenarios, PAFOT requires 16.05 fewer hours in total simulation time than AV-Fuzzer, and 21.79 hours less than Random.
    The experimental results show that compared with AV-Fuzzer and Random Fuzzer, PAFOT is able to find safety-critical scenarios in a shorter time. This shows that PAFOT is more efficient in exposing safety violations than the other testing techniques.
\end{custombox}

\section{Discussion and Future Work}

To increase the likelihood of perturbing the ADS in the testing scenarios, we introduce a novel position-based approach based on a 9-position grid virtually \textit{drawn} around the EV. This grid serves as a dynamic framework, allowing for precise adjustments to the positions NPCs under the control of a PID controller. By leveraging this position-based strategy, we systematically manipulate the spatial relationships of NPCs relative to the specified grid, thereby creating a dynamically challenging driving environment for the EV. The effectiveness of this position-based approach was evaluated by implementing it in the CARLA simulator. The results demonstrate that PAFOT, significantly outperforms baseline techniques in terms of generating a higher number of safety-critical scenarios within a shorter time frame. This showcases the potential of the proposed position-based methodology in systematically introducing perturbations that challenge the ADS in diverse and realistic driving conditions, showcasing its potential as an advanced and efficient technique for safety testing in autonomous driving scenarios.

At present, our approach is operational within the CARLA simulator and evaluated using the simulator’s built-in traffic manager ADS. CARLA’s traffic manager is designed with methods and modules that industrial-grade ADSs also utilise, including critical functionalities such as localisation, motion planning, vehicle tracking and sensing. This choice of simulator and ADS integration allowed us to conduct testing in a controlled yet realistic environment, providing valuable insights into the effectiveness and efficiency of our proposed technique. To extend to scope and assess the generalisability of our approach, future work is planned to evaluate the performance of PAFOT on established industrial-grade ADS platforms, such as Apollo~\cite{apollo} or Autoware~\cite{noauthor_autoware_2023}. By subjecting PAFOT to the complexity of industrial-grade ADSs, we aim to test and improve the robustness and real-world applicability of our proposed safety testing methodology.

In its current iteration, PAFOT operates on a single-objective fitness function to guide the search process. This approach has proven to be effective, particularly given the fitness criteria employed, (METTC, MD SD, and ET). However, it is of our interest to expand the guiding factors employed within PAFOT's optimisation process. Specifically, we are considering incorporating additional criteria, such as the diversity of generated scenarios, to improve the robustness and comprehensiveness of the testing technique. Following this objective, a shift towards utilising multi-objective GAs, such as NSGA-II, is planned. With this change we aim to incorporate the benefits of multi-objective optimisation, offering a more robust approach to generate scenarios with PAFOT. By including different and additional guiding factors, our aim is to further improve the effectiveness of PAFOT in finding safety-critical scenarios for AVs. 

\section{Related Work}

This section is dedicated to a comprehensive exploration of the existing research landscape within the domain of simulated test generation. By looking into prior research endeavours, we aim to clarify the current state of the field, identifying key methodologies and approaches that have shaped the landscape of simulated testing. Furthermore, this section serves to establish a clear demarcation between the PAFOT and previous works, exposing the distinctive contributions and advancements that our approach introduces to the domain. The literature review will involve a critical examination of various methodologies employed in simulated test generation, emphasising the distinctions and complexities inherent in these approaches. By drawing comparisons and contrasts, we seek to highlight the novel aspects and innovative features embedded in our proposed technique. This evaluative process aims to position our approach within the broader context of simulated testing, showcasing its potential to contribute significantly to the evolution and refinement of safety testing methodologies for ADSs.

A survey conducted by Lou et al. involving developers and testers in the Autonomous Vehicle (AV) industry underscores the crucial necessity of generating potential corner cases and unforeseen driving scenarios in AV testing~\cite{lou2022testing}. Therefore, various techniques have been put forward to produce safety-critical test scenarios for AVs~\cite{chandrasekaran2021combinatorial, gambi2019generating, han2020metamorphic, kluck2018using, li2020ontology, tao2019industrial, tian2018deeptest, zhang2018deeproad, zhou2020deepbillboard, tian2022generating, abdessalem2018testing, ben2016testing, calo2020generating, cheng2023behavexplor, yang2023lawbreaker, tang2023evoscenario, zhang2023testingads, zhou2023collver}.

Chandrasekaran et al. advocate a combinatorial testing approach to generate images for testing Deep Neural Network (DNN) models utilised in AVs~\cite{chandrasekaran2021combinatorial}. AC3R employs natural language processing and a domain-specific ontology to extract car crash driving scenarios from police reports~\cite{gambi2019generating}. M-CPS generates simulated scenarios by leveraging information from real accident images and videos~\cite{zhang2023building}. Kluck et al. utilise domain ontologies for providing environmental models and employ combinatorial testing to obtain critical scenarios~\cite{kluck2018using}. Zhong et al. propose a controllable traffic generation
method utilising broadcast modelling and Signal Temporal Logic (STL)~\cite{zhong2023guided}. This method facilitates the creation of lifelike traffic trajectories that emulate real-world traffic and adhere to specific objectives outlined by STL formulas. Metamorphic relations are employed in~\cite{zhang2018deeproad, tian2018deeptest, han2020metamorphic, zhou2019metamorphic} to test the behaviour of an Automated Driving System (ADS). Zhou et al. generate safety-critical scenarios by introducing adversarial perturbations to real-world billboards~\cite{zhou2020deepbillboard}. Tian et al. extract behaviour patterns leading to collisions from naturalistic driving datasets.~\cite{tian2022generating}. Stocco et al. utilise a Variational Autoencoder to distinguish between normal and anomalous behaviour of AV while regulating the detection's false positive rate~\cite{stocco2020misbehaviour, stocco2022confidence}. Biagiola and Tonella propose a test generation strategy by mutating the initial state conditions of EVs in fault-free scenarios and identifying challenging driving conditions at the behavioural boundary of these scenarios~\cite{biagiola2023boundary}. AutoFuzz uses Neural Networks to guide an evolutionary search over inputs of the autonomous vehicle API~\cite{zhong2022neural}. This technique entails training the network to predict whether new seeds will result in unique traffic violations, with the most promising seeds undergoing mutation to generate new adversarial inputs. BehAVExplor ~\cite{cheng2023behavexplor} presents a behaviour-guided fuzzer to explore different behviours of the ADS under test. LawBreaker ~\cite{yang2023lawbreaker} introduces a framework to test AVs utilising real-world traffic laws and find violations of traffic laws. EvoScenario ~\cite{tang2023evoscenario} presents a framework for testing AVs using road structures and segments using a multi-objective optimisation. Zhang et al. ~\cite{zhang2023testingads} propose ABLE, a testing approach to break traffic laws efficiently by dynamically updating testing objectives. Zhou et al. ~\cite{zhou2023collver} present CollVer, a testing framework to generate testing scenarios and judge whether the accident can be attributed to the ADS under test.

Search-based methods prove to be particularly effective for system testing in intricate problems such as Autonomous Vehicles (AVs)\cite{zeller2017search}. SAMOTA\cite{haq2022efficient} integrates surrogate-assisted optimisation and many-objective search to efficiently and effectively generate test scenarios. In the work by Abdessalem et al.\cite{abdessalem2018testing}, a multi-objective evolutionary algorithm is utilised to identify system-level failures in AVs caused by feature interactions. \cite{ben2016testing} involves using surrogate models with a multi-objective search to test the pedestrian detection system of an ADS. AsFault~\cite{gambi2019automatically} employs a genetic algorithm to explore road features that would push an AV to deviate from the lane centre. Likewise, tools released within the context of the Search-Based Software Testing Challenge (SBST)~\cite{panichella2021sbst, gambi2022sbst} produce intricate road networks for the virtual assessment of an automated lane-keeping system, such as GABezier~\cite{kluck2021gabezier}, Frenetic~\cite{frenetic}, WOGAN~\cite{peltomaki2022wogan}, and Deeper~\cite{moghadam2021deeper}. 

AV-Fuzzer~\cite{av-fuzzer}, MOSAT~\cite{tian2022mosat}, DoppelTest~\cite{huai2023doppelganger}, NADE~\cite{feng2021intelligent}, and ATLAS~\cite{Tang2021} propose test generation strategies to manipulate the manoeuvres of other vehicles (Non-playable Characters -- NPCs) on the road, introducing more challenging driving scenarios for AVs. Our work is closely related to these studies, as we also focus on generating challenging driving scenarios by searching for the challenging manoeuvres of NPCs. DoppelTest focuses on solving the problem of whether the collision is caused by a bug in the AV or due to some unavoidable situation like an NPC violating traffic rules and intentionally trying to hit the AV. NADE generates safety-critical scenarios by training the NPCs when to perform what adversarial manoeuvre, thus providing an intelligent environment for driving intelligence testing. AV-Fuzzer~\cite{av-fuzzer} employs a genetic algorithm (GA) integrated with fuzzing to identify and simulate challenging manoeuvres of NPCs, thereby inducing critical driving conditions for AVs. 
MOSAT~\cite{tian2022mosat} utilises genes to represent fundamental driving manoeuvres and employs a multi-objective genetic algorithm to explore diverse and adversarial test scenarios. AV-Fuzzer and MOSAT frequently explore driving operations that are situated at a considerable distance from the AV. As a consequence, a considerable portion of the optimization process for these techniques is dedicated to identifying driving patterns that would effectively draw traffic actors closer to the AV. In contrast, our proposed technique strategically models the operations of NPCs within a predefined 9-position grid around the AV. This deliberate choice streamlines the optimization process by removing the need to expend significant effort in the search for driving operations needed to bring closer the NPCs and the AV. By predefining and structuring the spatial relationship between NPCs and the AV, our approach not only reduces the computational complexity of the optimization task but also allows for a more targeted and efficient exploration of safety-critical scenarios, ultimately contributing to the efficacy and speed of the overall testing process.

\section{Threats to Validity}
A potential threat to the validity of our proposed approach revolves around the generation of realistic scenarios. To mitigate this thread, we have chosen to demonstrate the implementation of our technique within the CARLA simulator. CARLA, an open-source and widely adopted simulator in the domain of AV testing research, provides a realistic and dynamic environment for assessing the performance of AVs. The selection of CARLA not only establishes a solid foundation for showcasing the capabilities of our proposed technique but also aligns with the prevailing standards within the research community. Furthermore, it is important to emphasize that the versatility of PAFOT extends beyond the limits of CARLA. Our technique is designed to seamlessly integrate into any high-fidelity simulator that allows the representation of multiple NPCs and facilitates the manipulation of their actions through an API or other programmatic interfaces. This adaptability underscores the broader applicability of PAFOT, allowing researchers and practitioners to leverage its capabilities within diverse simulation environments, thus ensuring the generalisability and robustness of our proposed approach.

Another potential threat to the validity of our study can occur from the implementation of the baseline technique, AV-Fuzzer. Originally designed for the SVL simulator, this technique had to undergo a necessary re-implementation to suit the CARLA Simulator, considering the discontinuation of SVL. Such a transition introduces the possibility of discrepancies in the results due to variations in simulator characteristics and implementation details. To mitigate this threat, rigorous efforts were undertaken to ensure the fidelity of our re-implementation. Extensive testing was conducted, involving a meticulous comparison of our code against the original paper detailing AV-Fuzzer. This thorough validation process aimed to confirm the alignment of the results form our re-implementation with the outcomes reported in the original studies. 

The adoption of an \textit{elitist} selection strategy within our GA introduces a susceptibility to early convergence towards sub-optimal solutions. Recognising the potential ramifications of this tendency, and to safeguard against premature convergence that might compromise the optimality of the solutions, we strategically incorporated a restart mechanism, detailed in section~\ref{subsec:restart}. The restart mechanism serves as an intervention to counteract stagnation in the evolutionary process. When the GA encounters a notable lack of progress, in terms of fitness improvement across multiple generations, the restart mechanism is triggered. This mechanism generates new starting seeds for evolution, effectively redirecting the search towards unexplored regions of the search space. By implementing this restart mechanism, we mitigate the risk of entrapment in local sub-optimal solutions and promote a more thorough exploration of the search space. This dynamic approach contributes to the overall robustness and efficacy of our GA, enhancing its capacity to uncover diverse and potentially optimal solutions in the search for safety-critical scenarios.

\section{Conclusion}
We propose PAFOT, a search-based testing framework for testing Autonomous Vehicles by generating adversarial test scenarios to expose safety violations of ADSs. We proposed a 9-position grid system and introduced Position-Instructions for modifying the driving patterns and behaviours of third-party driving actors in test scenarios. The experimental results on the CARLA Simulator show that PAFOT is able to find more safety violations and safety-critical scenarios in less time when compared to existing testing techniques.

\section{Data Availability}
The code and instructions to replicate the implementation of PAFOT can be found on \url{https://github.com/pafotpafot/pafot}.

\bibliographystyle{ACM-Reference-Format}
\bibliography{references}

\end{document}